\documentclass[12pt]{article}
\newcommand{\xxx}[1]{ [#1]}
\newcommand{\mysection}[1]{\section{#1}
   \hspace{0.8cm}\setcounter{equation}{0}}

\newlength{\dummysp}
\newcommand{\beq}{\begin{equation}}
\newcommand{\eeq}{\end{equation}}
\newcommand{\ben}{\begin{enumerate}}
\newcommand{\een}{\end{enumerate}}

\newcommand{\tr}{\mathop{{\hbox{{\rm tr}} }}\nolimits}
\newcommand{\mtxt}[1]{\mathop{\hbox{{\small #1}}}\nolimits}

\newcommand{\half}{{1 \over 2}}

\newcommand{\beqa}{\begin{eqnarray}}
\newcommand{\eeqa}{\end{eqnarray}}
\newcommand{\nnn}{ \nonumber \\ }
\newcommand{\mod}{{\; \mtxt{{mod}} \; }}
\newcommand{\p}{{\partial}}
\newcommand{\Lcal}{{\cal L}}

\newcommand{\s}{{\sigma}}

\newcommand{\vev}[1]{{\langle #1 \rangle}}

\newcommand{\ord}[1]{{{\cal O}(#1)}}

\newcommand{\gappeq}{\mathrel{\rlap {\raise.5ex\hbox{$>$}}
{\lower.5ex\hbox{$\sim$}}}}
\newcommand{\lappeq}{\mathrel{\rlap{\raise.5ex\hbox{$<$}}
{\lower.5ex\hbox{$\sim$}}}}
\newcommand{\myref}[1]{(\ref{#1})}

\newcommand{\ux}{$U(1)_X$}

\newcommand{\LamH}{\Lambda_H}
\newcommand{\LamX}{\Lambda_X}
\newcommand{\hc}{{\rm h.c.}}
\newcommand{\yoa}{|X^\alpha|}
\newcommand{\yta}{|Y^\alpha|}
\newcommand{\yoj}{|X^j|}
\newcommand{\ytj}{|Y^j|}
\newcommand{\yok}{|X^k|}
\newcommand{\ytk}{|Y^k|}
\newcommand{\lra}{\leftrightarrow}
\newcommand{\A}{{\cal A}}
\newcommand{\B}{{\cal B}}
\def\[{\left[}
\def\]{\right]}
\def\({\left(}
\def\){\right)}

\begin{document}

\begin{titlepage}

\def\thefootnote{\fnsymbol{footnote}}

\baselineskip=14pt

\hfill    LBNL-51212

\hfill    UCB-PTH-02/33

\hfill    hep-ph/0208004

\hfill    Sept.~27, 2002

\vspace{10pt}

\begin{center}
{ \bf \Large D-Moduli Stabilization}\footnote{Based
on a talk given at the Superstring Phenomenology Conference,
University of Oxford, England, July 7, 2002.
Extended version of what is to appear in the conference
proceedings.  Based on work done in collaboration
with Mary K. Gaillard.}
\end{center}

\begin{center}
{\sl Joel Giedt}\footnote{E-Mail: {\tt JTGiedt@lbl.gov}}
\end{center}

\begin{center}
{\it Department of Physics, University of California, \\
and Theoretical Physics Group, 50A-5101, \\
Lawrence Berkeley National Laboratory, Berkeley,
CA 94720 USA.}\footnote{This work was supported in part by the
Director, Office of Science, Office of High Energy and Nuclear
Physics, Division of High Energy Physics of the U.S. Department of
Energy under Contract DE-AC03-76SF00098 and in part by the National
Science Foundation under grant PHY-00-98840.}
\end{center}

\begin{center}
{\bf Abstract}
\end{center}

The matter sector
of four-dimensional effective supergravity
models obtained from the weakly coupled
heterotic string contains many moduli.
In particular, flat directions of the D-term
part of the scalar potential in the presence
of an anomalous $U(1)$ give rise to massless
chiral multiplets which have been referred to
elsewhere as {\it D-moduli.}
The stabilization of these moduli is necessary for
the determination of the large vacuum expectation
values of complex scalar fields induced by the
corresponding Fayet-Illiopoulos term.  This
stabilization is of phenomenological importance since these
background values determine the effective theory below
the scale of the anomalous $U(1)$ symmetry
breaking.  In some simple models we illustrate
the stabilization of these moduli
due to the nonperturbative dynamics
associated with gaugino condensation
in a hidden sector.  We find that background field
configurations which are stable above the condensation
scale no longer represent global minima once
dynamical supersymmetry breaking occurs.
The implications for low energy models based
on promising ``flat'' directions are
discussed.

\vfill

\end{titlepage}

\setcounter{page}{1}
\def\thefootnote{\arabic{footnote}}
\setcounter{footnote}{0}

\baselineskip=14pt

\mysection{Introduction}
Cancellation of the trace anomaly associated with an anomalous
\ux\ by the GS mechanism \cite{GS} leads to an
FI term \cite{UXR} for the D-term of \ux:
$$
D_X = \sum_i K_i q_i^X \phi^i + \xi, \qquad
\xi = {g_H^2 \tr Q_X \over 192 \pi^2} m_P^2 .
$$
$K_i = \p K/\p \phi^i$ is the field
derivative of the K\"ahler potential,
$g_H$ is the unified coupling at the
string scale $\LamH$, the charge
of the scalar field $\phi^i$ is given by $q_i^X$  and $m_P = 1/\sqrt{8\pi G}
= 2.44 \times 10^{18}$ GeV is the reduced Planck mass.

Large matter field vacuum expectation
values ({\it vevs}) are generic in semi-realistic
string-inspired models because anomalous
\ux's are generic.
In \cite{GieB} it was shown that
the presence of a \ux\ factor in the gauge group $G$
is generic for semi-realistic $Z_3$ orbifold models.
For the class of standard-like orbifolds studied
there, only $7$ of $175$ models {\it did not} have a \ux.
In the semi-realistic free fermionic
models \cite{CFN} a \ux\ is also generic.

The $n$ fields $\phi^i$ which acquire nonvanishing
vevs $\vev{\phi^i} \equiv v^i$
will be referred to here as {\it X-Higgs} fields.
For canonical $K= \sum_i |\phi^i|^2$, we have 1 constraint:
\beq
\vev{D_X} = \sum_i q_i^X |v_i|^2 + \xi.
\label{ae9}
\eeq
These vevs are generically complex.
Indeed, \myref{ae9} is completely
``phase-blind.''
In addition, $\vev{D_X}=0$ only constrains the
magnitudes of the vevs to take values on a
$n$-dimensional hyperboloid.
The moduli which parameterize the unconstrained complex
phases and the location on the $n$-dimensional hyperboloid
are flat directions of the D-term part of the scalar potential and
are pseudo-Goldstone bosons which were termed {\it D-moduli}
in a previous work done in collaboration with Mary K.
Gaillard (MKG) \cite{GGA}.
The D-moduli correspond to the $U(r,n-r)$ invariance of
$\vev{D_X}=0$, where
$r$ is the number of fields with $q_i^X >0$.
Only one of these D-moduli chiral multiplets gets
``eaten'' by the \ux\ vector multiplet when it
gets massive.

This vacuum degeneracy is a generic
problem in supersymmetric models \cite{BDFS}.
Typically one chooses a flat direction which
gives rise to ``good'' phenomenology.  However
this is arbitrary and may not be consistent
with the dynamics employed to break supersymmetry.
An effective theory of supersymmetry breaking
can potentially lift most vacuum degeneracy.  Accounting
for effects of dynamical supersymmetry breaking,
{\it arbitrariness in the phenomenology due
to X-Higgs vevs can be removed.}

Here, we assume supersymmetry breaking
via gaugino condensation in a hidden sector.
Improvements
in our understanding of strongly coupled super-Yang-Mills
(embedded into string-derived effective supergravity) increase
the reliability of this approach toward solving
vacuum selection problems.  It is already known that
different string embeddings are related to each other
by X-Higgs vevs.
For example, in the work of Aldazabal et al.~\cite{sGUT}
$k=1$ constructions and $k=2$ constructions are
related at special values of X-Higgs vevs.
Thus an effective field theory approach
to dynamical vacuum selection can make modest progress
in the vacuum selection problem of string theory.

\mysection{Scope}
Efforts are underway in work with MKG to stabilize
these moduli in semi-realistic models
by including various terms in $V$
(intentionally) neglected in our earlier work \cite{GGA}.
MKG has found ways to:
consistently fix to unitary gauge (not an easy task
with dynamical string moduli);
preserve manifest target-space modular invariance;
effectively include tree-level exchange of
heavy multiplets in the
general case of many fields and $U(1)$'s.
The machinery is forthcoming \cite{GGB}.

Here I report only on stabilization in some rather
simple ``toy'' models.
I do not account for tree-level
exchange of the heavy fields.
I do not include compactification moduli
or target space modular invariance.
I oversimplify so as to isolate
the issue of D-moduli stabilization.
Fixing to unitary gauge is simple when string moduli
are treated as constant background fields.

\mysection{Background}
In \cite{GGA}, the scalar potential $V$
for SUGRA with a \ux\ was studied for vacuum configuations
satisfying
\beq
\vev{V} = \vev{\p V / \p \phi^i } = 0.
\label{hya}
\eeq
Supersymmetry breaking was characterized by
$$
{1 \over m_P^4} \vev{|W|^2} = |\delta|^2,
\qquad
{1 \over m_P^2} \vev{K_{i \bar \jmath} F^i \bar F^{\bar \jmath} }
= \alpha |\delta|^2 e^{\vev{K}/m_P^2} ,
\qquad
\alpha \sim \ord{1} .
$$
According to expectations, it was found that
\myref{hya} together with a reasonable
supersymmetry breaking scale $|\delta| \sim 1$ TeV requires
$$
\vev{D_X} \sim |\delta|^2 \ll |\xi| .
$$
We found that in the stable vacuum only fields
with the minimum charge $\min \{q_i \}$ can
get vevs.  One combination of fields from this
set gets eaten by the \ux\ multiplet while
the remainder are massless {\it after} supersymmetry
breaking!
For canonical K\"ahler potential
$$
\vev{D_X} = \sum_i q_i^X |v_i|^2 + \xi, \quad
q_i^X \sim 1 \quad
\Rightarrow \quad |v^i| \sim \sqrt{|\xi|} .
\nonumber
$$
Research in progress with MKG finds that
these order of magnitude relations
hold in cases more complicated
than those studied in our earlier work.
This is also in agreement with work of
Barreiro et al.~\cite{Bar}.

Based on $|v^i| \sim \sqrt{|\xi|}$, the \ux\
gauge symmetry breaking scale $\LamX$
may be defined as
$$
\LamX = \sqrt{|\xi|} .
$$
For the class of models studied in \cite{GieB} it was found
that for the 168 of 175 cases where $\xi \not= 0$,
$$
{g_H \over 8.00} \leq {\LamX \over m_P}
\leq {g_H \over 4.63} = {\LamH \over m_P} .
$$
where $\LamH \approx 0.216 \times g_H m_P$ is the
approximate string scale obtained by Ka\-plu\-nov\-sky~\cite{Kap}.
With $g_H \sim 1$ we have that $\LamX \sim 0.1 \times m_P$
is a generic prediction.
The result of this is that nonrenormalizable operators
should contribute {\it significantly} to the
(effective) Yukawa couplings of the lighter
quarks, since they are only down by
$(\LamX / m_P)^n \sim 10^{-n}, n>0$.
Operators with $1 \leq n \leq 4$
would typically be present.\footnote{It is worth noting,
however, that along certain flat directions in explicit
string constructions, effective mass operators for
light fields are forbidden by selection rules at
all nonrenormalizable orders of the
superpotential, as has been emphasized
recently in \cite{Mun01}.}
Given $\lambda_{u,d}/\lambda_t \sim 10^{-5}$ after
running to the high scale, it is difficult to believe that
nonrenomalizable operators would
not play a role, generically
speaking.
This serves as an example of how
stabilization of the D-moduli is a crucial
ingredient in predicting low-energy physics.

\mysection{Effective scalar potential}
This is a modification of the linear multiplet ($L$)
toy model considered in earlier work with MKG \cite{GGA}.
The desire is to lift
vacuum degeneracy by coupling D-moduli to matter condensates
of the hidden sector condensing group $G_C$.
Such couplings are expected\footnote{I thank Emilian Dudas
for pointing this out to us.} from the mixed trace anomaly
matching condition $\tr T^a T^a Q_X \not= 0,$
where $T^a$ is a generator of $G_C$.
We have
\beqa
K &=& k(L) + G(A,B,\Phi,\bar A,\bar B,\bar \Phi), \qquad
k(L) = \ln L + g(L), \nnn
G &=& \sum_i |A_i|^2 + \sum_i |B_i|^2 + \sum_i |\Phi_i|^2 .
\nonumber
\eeqa
The chiral superfields $\Phi_i$ are supposed to be
the X-Higgses.  We denote the scalar components $\phi_i$
and the corresponding vevs $v_i = \vev{\phi_i}$.
The chiral superfields $A_i$ and $B_i$
are supposed to be charged under an unbroken
factor of the low energy gauge group, such as $SU(3)_c$,
so that they are forbidden from acquiring vevs.

I add a term $\breve W$ to the superpotential so
that it now takes the form
\beqa
W(A,B,\Phi,\Pi) &=& \hat W(A,B,\Phi) + \breve W(\Phi,\Pi), \nnn
\hat W(A,B,\Phi) &=& \sum_{i,j,k} \lambda_{ijk} A_i B_j \Phi_k, \qquad
\breve W(\Phi,\Pi) = \sum_\alpha c_\alpha(\Phi) \Pi_\alpha .
\label{ce2}
\eeqa
Here, $\Pi_\alpha$ are hidden sector matter condensate
superfields.
The functional $c_\alpha(\Phi)$ is left
unspecified at this point.
We implement dynamical supersymmetry breaking
through a Veneziano-Yankielowicz-Taylor effective
Lagrangian \cite{VYT}, following Bin\'etruy, Gaillard
and Wu (BGW) \cite{BGW}:
$$
\Lcal_{{\rm VYT}} = \int {E \over 8R} U \[ b' \ln (e^{K/2} U)
+ \sum_\alpha b^\alpha \ln \Pi^\alpha \] + \hc
$$
The chiral superfield $U$ corresponds to the condensing
gaugino bilinear and the coefficients $b'$ and $b^\alpha$
are determined by anomaly matching.
We have no compactification moduli appearing, no threshold
corrections, and the only GS term is the one required
to cancel the \ux\ anomaly.
Following the BGW formulation
one obtains for the scalar potential
\beqa
V &=& \half \( {2 \ell \over 1+f(\ell)} \) \sum_a D_a D_a
+ (\ell g'(\ell) - 2) \left| {b' u \over 4} - e^{K/2} W \right|^2 \nnn
&& + \left|e^{K/2} (W_I + W G_I) - {b' u \over 4} G_I \right|^2 \nnn
&& + \( {1+ \ell g'(\ell) \over 16 \ell^2} \)
\[ (1+2 \ell b') |u|^2 - \ell e^{K/2} (W \bar u + \bar W u) \] .
\nonumber
\eeqa
Here, $\ell = L|, u=U|$ and the functional $f(\ell)$ is
closely related to the nonperturbative correction $g(\ell)$
to the dilaton K\"ahler potential.
Nevermind all the details in $V$; the point is that
in principle we can find the stable vacua.  What
remains is purely a technical challenge.

We restrict our attention to the case
where $D_X$ is the
only nonvanishing D-term, and $\vev{A_i}=\vev{B_i}=0$.
In this case $\vev{V}$ is,
after straightforward manipulations,
given by
\beqa
V&=& \half g_H^2 D_X^2 + \hat V, \nnn
\hat V & = & e^K \sigma^2
   \[ b_c^2 (v^2-2+\ell g')
   + \( { 1+\ell g' \over 2 \ell^2} \)
   (2+3 \ell b' + \ell b_c) \] ,
\label{owt}
\eeqa
where all quantities from here on out
are taken at their vevs and
\beqa
v&=&\[ \sum_i |v_i|^2 \]^{1/2}, \qquad
g_H^2 = {2 \ell \over 1 + f(\ell)}, \nnn
K &=& k(\ell) + v^2, \qquad
b_c = b' + \sum_\alpha b^\alpha, \nnn
D_X &=& \sum_i q_i |v_i|^2 + \xi, \qquad
\s = {b' \over 4} e^{-K/2} |u|, \nnn
\s & = & {1 \over 4}
   \exp \[ - {1 \over b_c g_H^2} -  {b' \over b_c} \]
   \prod_\alpha \left| {4 c_\alpha(v)
   \over  b^\alpha } \right|^{ b^\alpha / b_c} .
\nonumber
\eeqa
Note that $\s$ is essentially a reparameterization
of the gaugino condensate; i.e., it is the order
parameter for supersymmetry breaking.
From these expressions it is not hard to work out
$V_i = \p V/\p v_i$:
\beq
V_i  = \bar v_i \[ \A q_i + \B  \] + \mu_i \hat V ,
\label{ce8}
\eeq
where
\beq
\A \equiv g_H^2 D_X, \qquad
\B \equiv b_c^2 \s^2 e^K + \hat V, \qquad
\mu_i \equiv \sum_\alpha {b^\alpha \over b_c} {\p \over \p v_i}
\ln c_\alpha (v) .
\label{ce9}
\eeq

Note that the term $\mu_i \hat V$ in \myref{ce8}
was not present in our previous work.  Then for
the nonvanishing vevs we had only 1 constraint:
$\A q_i + \B = 0$ for the minimum charge $q_i = -q$.
However we now have the term $\mu_i \hat V$
due to the coupling $\breve W(\phi,\pi)$ and consequently
$n$ constraints on the $n$ fields getting vevs.
Thus we expect that
\myref{ce8} provides the necessary constraints
to lift the D-moduli flat directions,
barring flavor symmetries which might
lead to redundant equations.
In addition to the vanishing of \myref{ce8},
we also impose $V=0$.
Analysis of these two conditions,
keeping in mind $\s^2 \ll |\xi|$,
leads to the results:
\beqa
|v_i| &\sim &
\left\{
\begin{array}{l}
\LamX = \sqrt{|\xi|} \qquad q_i = -q , \\
\s \qquad q_i \not= -q ;
\end{array}
\right. \nnn
q &=& - \min \{ q_i \}.
\label{vsz}
\eeqa
Thus we get a {\it considerable} vacuum selection:
only fields with $q_i = -q$ can get large vevs;
the remainder get vevs of order the supersymmetry
breaking scale.
In many cases this may be sufficient to {\it rule out}
flat directions\footnote{More precisely,
directions which were flat in the absence of supersymmetry breaking.}
which were assumed for phenomenological
reasons.  The pleasing feature of this result is that it
does not require a detailed knowledge of the
form of $\breve W(\phi,\pi)$.

Note that we have only considered
the case with X-Higgses charged solely under \ux.
The analogous vacuum selection which occurs for
the more general case of X-Higgses charged under
several factors of the gauge group will be
considered elsewhere~\cite{GGB}.

Notice that all of the quantities in \myref{ce8} are real except
$\bar v_i$ and $\mu_i$.
From \myref{ce8} we see that the phase of $v_i$ will
be related to the phase of $\mu_i$.
More precisely,
$$
\arg v_i = - \arg \mu_i \mod \pi .
$$

We next suppose in \myref{ce2}
\beq
c_\alpha(v) = \sum_A c_{\alpha A}(v), \qquad
c_{\alpha A}(v) = \lambda_{\alpha A} \prod_i (v_i)^{p_{iA}^\alpha} .
\label{ce14}
\eeq
Then it is easy to check that \myref{ce9}
yields
$$
v_i \mu_i = \sum_\alpha {b^\alpha \over b_c}
{ \sum_A p_{iA}^\alpha c_{\alpha A}(v) \over
\sum_A c_{\alpha A}(v) } .
$$
Consequently we can rewrite the minimization constraint
which follows from \myref{ce8} as
\beq
0 = |v_i|^2 \[ \A q_i + \B \]
+ \hat V \sum_\alpha {b^\alpha \over b_c}
{ \sum_A p_{iA}^\alpha c_{\alpha A}(v) \over
\sum_A c_{\alpha A}(v) } .
\label{ce16}
\eeq
In the case where the sum in \myref{ce14} has only a single
term, the $c_{\alpha A}(v)$ cancel in \myref{ce16} and
no phase constraints exist.
Thus, a non-monomial polynomial assumption
for $c_\alpha(v)$ is required for phase stabilization.

\mysection{A simple example}
As an example, we review the simple case
considered previously in \cite{GGC},
the case of only two fields
$\phi^1,\phi^2$ of charges $q_1=q_2 \equiv -q$ and a single
matter condensate field $\pi$ with superpotential coupling
$$
\breve W(\phi,\pi) = c(\phi) \pi, \qquad
c(\phi) = \lambda_1 \phi_1 + \lambda_2 \phi_2 .
$$
We define
$$
v_1 = e^{i\varphi_1} v \cos \eta, \qquad
v_2 = e^{i\varphi_2} v \sin \eta .
$$
$\varphi_1-\varphi_2$ is the phase we would like to stabilize and
$\eta$ is the mixing angle to the mass eigenstate basis which
we would also like to stabilize.  These are the D-moduli.
The scalar modes corresponding to $v$ and $\varphi_1 + \varphi_2$
are eaten by the \ux\ vector multiplet.

It is not hard to check that \myref{ce8} gives
$$
0 = b_c ( \lambda_1 |v_1|^2 + \lambda_2 \bar v_1 v_2)
(\B- q \A) + b^\alpha \lambda_1 \hat V ,
$$
and a similar equation with $1 \lra 2$, and then 2 conjugate
equations.
Manipulations on these four equations lead simply to
$$
{v_1 \bar v_2 \over \bar v_1 v_2} =
{\bar \lambda_1 \lambda_2 \over \lambda_1 \bar \lambda_2}
\Rightarrow
\varphi_1-\varphi_2 =
\arg({\lambda_2 \over \lambda_1}) \mod \pi .
$$
It is also straightforward to check
$$
\sin^2 \eta = {b^\alpha \hat V (|\lambda_1|^2 - |\lambda_2|^2)
+ b_c v^2 |\lambda_1|^2 (\B-q\A)
\over 2 b_c v^2 |\lambda_2|^2
(\B-q\A)} .
$$
Thus the D-moduli are stabilized and the phase and mixing
are determined.

\mysection{A less simple example}
We have fields $S,X^i,Y^i$ which
are X-Higgses evaluated at their vevs.
We assume charges $Q_X(S) = q_S$,
$Q_X(X^i) = q_X$, $Q_X(Y^i) = q_Y$,
($\forall \; i=1,2,3$) satisfying
$q_S = \min (q_S,q_X,q_Y) = -2 q_X = -2 q_Y$.
It is convenient to define
$w^\alpha(X,Y) = X^j Y^k + X^k Y^j$
where $\alpha \not= j \not= k \not= \alpha$
everywhere here and below.
We assume matter condensates $\Pi_\alpha \; (\alpha=1,2,3)$ in the hidden
sector and we impose string-inspired discrete symmetries
on the couplings such that
$$
\breve W = \sum_{\alpha=1}^3 c_\alpha(S,X,Y) \Pi_\alpha, \qquad
c_\alpha(S,X,Y) = \lambda S w^\alpha(X,Y).
$$

We examine generic points in moduli space, for which
none of the fields $S,X^i,Y^i$ vanish and
$\A q_S + \B \not=0$, $\A q_X + \B \not=0$, $\A q_Y + \B \not=0$.
Applying the conditions of minimization and cancellation
of the cosmological constant, we have
$$
|S|^2 = {(b' - b_c) \hat V \over b_c (\A q_S + \B)}
$$
and the constraints (together with conjugates)
\beqa
0 &=& b_c (\A q_X + \B) w^j w^k |X^\alpha|^2
+ b^\alpha \hat V ( w^j X^\alpha Y^j + w^k X^\alpha Y^k ),
\nnn
0 &=& b_c (\A q_Y + \B) w^j w^k |Y^\alpha|^2
+ b^\alpha \hat V ( w^j Y^\alpha X^j + w^k Y^\alpha X^k ).
\nonumber
\eeqa

By judiciously combining these equations it is not difficult to
show that
$$
0 = b_c(\A q_X+\B)|X^\alpha|^2+b_c(\A q_Y+\B)|Y^\alpha|^2
+2b^\alpha\hat V.
$$
The 3 scale moduli $\eta_\alpha$ which parameterize solutions
to these equations are defined by ($\hat V < 0$)
\beqa
b_c(\A q_X+\B)|X^\alpha|^2 &\equiv& -2b^\alpha\hat V\cos^2\eta_\alpha, \nnn
b_c(\A q_Y+\B)|Y^\alpha|^2 &\equiv& -2b^\alpha\hat V\sin^2\eta_\alpha.
\label{vee}
\eeqa
Note that $\eta_\alpha$ correspond to 3 real scalars
which remain {\it massless} even after supersymmetry breaking
and the superpotential interactions $c_\alpha$ are included.
In the model studied here $q_S$ is the minimum charge
so according to \myref{vsz} it is this field
which mostly cancels the FI term.
On the other hand, $q_S = -2 q_X = -2 q_Y$.  Then
it is easy to show that minimization subject
to cancellation of the cosmological constant yields
$$
\A q_S+\B= \ord{\s^4}, \quad \A q_X+\B = \A q_Y+\B
= {3 \over 2} \B + \ord{\s^4}.
$$
Noting $\B = \ord{\s^2}$ and $\hat V = \ord{\s^4}$
we have from \myref{vee}
$$
\yoa^2 = {4 b^\alpha |\hat V| \over 3 b_c \B} \cos^2 \eta_\alpha
+ \ord{\s^4} = \ord{\s^2},
$$
$$
\yta^2 = {4 b^\alpha |\hat V| \over 3 b_c \B} \sin^2 \eta_\alpha
+ \ord{\s^4} = \ord{\s^2},
$$
in agreement with \myref{vsz}.

After some manipulation the constraints on complex
phases are found to be
\beqa
0 &=& 2 \yoa^2 \ytj \ytk \sin^2 \eta_\alpha
+ 2 \yta^2 \yoj \yok \cos^2 \eta_\alpha \cdot e^{i \beta_3^\alpha}
\nnn && + 2 (2 - \cos^2 \eta_\alpha) \yoa \yta
\( \yoj \ytk e^{i \beta_2^\alpha}
+ \ytj \yok e^{i \beta_4^\alpha} \)
\nonumber
\eeqa
where
$$
\beta_3^\alpha = 2(\phi_Y^\alpha-\phi_X^\alpha)
- (\phi_Y^j-\phi_X^j) - (\phi_Y^j-\phi_X^j),
$$
$$
\beta_2^\alpha = (\phi_Y^\alpha-\phi_X^\alpha)
- (\phi_Y^j-\phi_X^j), \qquad
\beta_4^\alpha = (\phi_Y^\alpha-\phi_X^\alpha)
- (\phi_Y^k-\phi_X^k),
$$
and $\phi_X^i = \arg X^i, \phi_Y^i = \arg Y^i$.
These 3 constraints on the 3 independent phases
$(\phi_Y^i - \phi_X^i)$ fix these pseudoscalar
D-moduli.
On the other hand, we have 3 orthogonal phases
$(\phi_Y^i + \phi_X^i)$ which do not get fixed,
corresponding to 3 massless pseudoscalar moduli.
The phase of $S$ also was not fixed by the
minimization conditions.
One linear combination of these 4 pseudoscalar moduli
is eaten by the \ux\ vector boson when it becomes
massive.
Thus, we are left with 3 pseudoscalar
D-moduli which remain massless after taking into
account supersymmetry breaking and the superpotential
interaction $c_\alpha$.
These pair up with the 3 real massless scalars
corresponding to $\eta_\alpha$ to give three
complex massless scalars.

\mysection{Conclusions}
Presumably, loop effects and additional terms added
to the superpotential would stabilize the remaining
moduli in the last example.
I have (here) only examined models with simplified
X-Higgs content and couplings relative to semi-realistic string
models.  Already the analysis is tedious.
To study the vacuum generally would be
rather involved and does not make much sense to do
unless the model is exceptionally promising.
In semi-realistic cases the litany of nonrenormalizable
superpotential interactions which might play a significant
role in D-moduli stabilization poses a technical challenge
for understanding the structure of the vacuum.
It is impossible to perform a systematic
numerical scan of the parameter space
spanned by $\ord{50}$ independent vevs.
However, such technical challenges have been
overcome in other subfields of physics, such
as nuclear, atomic and lattice gauge, through
{\it semi-analytic} techniques and {\it importance
sampling.}
For instance, a Metropolis algorithm which minimizes $V$,
or other advanced techniques for minimization of a
nonlinear function of many variables, may give us
a handle on global minima.
Local minima identified by such techniques may
represent metastable vacua with interesting
cosmological consequences.
Once minima are identified numerically, one could perhaps
expand about these minima analytically and check
for moduli; i.e., remaining flat directions.

\vspace{0.20in}

\noindent {\bf \Large Acknowledgments}

\vspace{5pt}

\noindent 
Special thanks are extended to
my collaborator in this research, Mary K. Gaillard.
The author also thanks the organizers of the
Superstring Phenomenology Conference for their hard
work and hospitality.  
This work was supported in part by the Director, Office of Science,
Office of High Energy and Nuclear Physics, Division of High Energy
Physics of the U.S. Department of Energy under Contract
DE-AC03-76SF00098 and in part by the National Science Foundation under
grant PHY-0098840.

\end{document}